\documentclass[aps,twocolumn]{revtex4}
\usepackage{graphicx}
\usepackage{epsfig}
\usepackage{amsmath,amssymb,color}
\usepackage{epstopdf}
\parskip=\medskipamount

\newcommand{\eq}[1]{(\ref{#1})}
\newcommand{\fig}[1]{Fig.\ref{#1}}

\newcommand{\be}{\begin{equation}}
\newcommand{\ee}{\end{equation}}

\begin{document}

\title{Eigenvalue tunnelling and decay of quenched random network}

\author{V. Avetisov$^{1,8}$, M. Hovhannisyan$^2$, A. Gorsky$^{3,4}$, S. Nechaev$^{5,6}$,
M. Tamm$^{7,8}$, O. Valba$^8$}

\address{$^{1}$N.N. Semenov Institute of Chemical Physics of the Russian Academy
of Sciences, 119991, Moscow, Russia \\ $^2$Chair of Programming and Information Technologies,
Yerevan State University, Yerevan, Armenia \\ $^3$Institute of Information Transmission Problems of
the Russian Academy of Sciences, Moscow, Russia, \\ $^4$Moscow Institute of Physics and Technology,
Dolgoprudny 141700, Russia \\
%$^5$Universit\'e Paris-Sud/CNRS, LPTMS, UMR8626, B\^at. 100, 91405 Orsay, France, \\
$^5$Poncelet Laboratory, CNRS (UMI2615), Independent University of Moscow, Moscow, Russia \\
$^6$P.N. Lebedev Physical Institute of the Russian Academy of Sciences, 119991, Moscow, Russia \\
$^7$Physics Department, Moscow State University, 119992, Moscow, Russia, \\ $^8$Department of
Applied Mathematics, National Research University Higher School of Economics, 101000, Moscow,
Russia.}

\begin{abstract}

We consider the canonical ensemble of $N$-vertex Erd\H{o}s-R\'enyi (ER) random topological graphs
with quenched vertex degree, and with fugacity $\mu$ for each closed triple of bonds. We claim
complete defragmentation of large-$N$ graphs into the collection of $[p^{-1}]$ almost full
subgraphs (cliques) above critical fugacity, $\mu_c$, where $p$ is the ER bond formation
probability. Evolution of the spectral density, $\rho(\lambda)$, of the adjacency matrix with
increasing $\mu$ leads to the formation of two-zonal support for $\mu>\mu_c$. Eigenvalue tunnelling from
one (central) zone to the other means formation of a new clique in the defragmentation process. The
adjacency matrix of the ground state of a network has the block-diagonal form where number of
vertices in blocks fluctuate around the mean value $Np$. The spectral density of the whole network
in this regime has triangular shape. We interpret the phenomena from the viewpoint of the
conventional random matrix model and speculate about possible physical applications.

\end{abstract}

\maketitle

%%%%%%%%%%%%%%%%%%%%%% Introduction %%%%%%%%%%%%%%%%%%%%%%%%%%%%%

Investigation of critical and collective effects in graphs and networks has becoming a new rapidly
developing interdisciplinary area, with diverse applications and variety of questions to be
asked, see \cite{dorog} for review. Ensembles of random Erd\H{o}s-R\'enyi topological graphs
(networks) provide an efficient laboratory for testing collective phenomena in statistical physics
of complex systems, being also tightly linked to conventional random matrix theory. Besides
investigating typical statistical properties of networks, like vertex degree distribution,
clustering coefficients, "small world" structure etc, last two decades have been marked by rapidly
growing interest in more refined graph characteristics, such as distribution of small
subgraphs involving triads of vertices.

Triadic interactions, being the simplest interactions beyond the free-field theory, play crucial
role in the network statistics. Presence of such interactions is responsible for emergence of phase
transitions in complex distributed systems. First example of a phase transition in random networks,
known as Strauss clustering \cite{strauss}, has been treated by the Random Matrix Theory (RMT) in
\cite{burda}. It was argued that, when the increasing fugacity, $\mu$, the system develops two
phases with essentially different triad concentrations. At large $\mu$ the system falls into the
Strauss phase with the single clique of nodes. The condensation of triads is a non-perturbative
phenomenon identified in \cite{newman} with the 1st order phase transition in the framework of
mean-field cavity-like approach.

Similar critical behavior was found in \cite{valba} for the vertex-degree-conserved ER graphs. It
was demonstrated in the framework of the mean-field approach that the phase transition takes place
in this case as well. The hysteresis for dependence of the triad concentration on the fugacity,
$\mu$ also has been observed in \cite{valba}. For bi-color networks with conserved vertex degree a
new  phenomena of a wide plateau formation in concentration of black-white bonds as a function of
the fugacity of unicolor triples of bonds has been found in \cite{color}.

Note that all these models are essentially athermic: in the absence of an external field, the
partition function of the network is a purely combinatorial object with no interactions and no
temperature dependence, and its evolution can be regarded as a Langevin dynamics in the stochastic
quantization framework. The solution to the corresponding Fokker-Planck equation at the infinite
stochastic time yields the exact quantum ground state of the model.

Here we provide deeper insight into the structure of phase transition in non-directed
vertex-degree-conserved Erd\H{o}s-R\'enyi random graphs, with the number of fully connected triples
of vertices, $n_{\triangle}$, controlled by $\mu$. The numerical simulations possess a number of
striking phenomena: (i) above some critical value, $\mu_c$, the network splits into the
\emph{maximally} possible number of clusters, identified as cliques, (ii) the number of cliques for
large graphs is fixed \emph{exclusively} by the average vertex degree at the network preparation,
$p$, (iii) above $\mu_c$ the spectral density of the whole network has the triangular shape,
typically observed for scale-free networks. This behavior differs significantly from the
unconstrained Strauss model, where above the transition point the single maximally connected clique
is formed.

Qualitatively the formation of the Strauss condensate in the unconstrained network can be
understood as follows. For $\mu=0$ the system lives in the largest entropic basin corresponding to
some equilibrium distribution of triads. As $\mu$ is increasing, the triad distribution gets
gradually more skewed. In the limit $\mu \to \infty$, the entropic effects become irrelevant, and
the network approaches the state with the largest energy, $\mu n_{\triangle}$. Depending on the
shape of the entropy, the dependence $n_{\triangle}(\mu)$ can be either a smooth function, or can
undergo abrupt jumps at some particular values of $\mu_c$ typical for first-order phase
transitions. In contrast, in the vertex-degree-conserved model the constraints prohibit the
formation of a large single clique since the constraints prevent of complete mixing. Therefore the
system does its best and splits into the maximally possible number of allowed cliques, which is the
true ground state for the network with quenched vertex degree. The same decay occurs for so-called "regular
random networks" (which have the one and the same degree $Np$ in all nodes).

Any topological graph or collection of graphs can be encoded by the adjacency matrix, $A$
\cite{dorog1}. In the quenched model, within the transition region, isolated eigenvalues of $A$
form the second zone in the spectrum and correspond one-by-one to clusters in the large network
(see \cite{newman2,newman3,newman4} for general description). Above the transition point, the
spectral density (SD), $\rho(\lambda)$, of the adjacency matrix of each clique (almost fully
connected subgraph) is the same as the spectral density of the sparse matrix, and has Lifshitz
tails typical for 1D Anderson localization, as discussed in \cite{av-kr-nech}. The spectral density
of the whole network has a triangle-like shape typically seen in scale-free networks. Note, however
that in our system the set of vertex degrees is quenched at the preparation, and has the Poisson
distribution.

Found behavior has many parallels in the random matrix model. The effective potential of the
Strauss model involves the quadratic and cubic terms \cite{burda}. The conventional random matrix
model attributes the particular Riemann surface to the effective potential due to the loop
equations. In the cubic case it yields the genus one Riemann surface for the SD, therefore
emergence of two zones in the support of $\rho(\lambda)$ of the adjacency matrix, is natural. The
distribution of eigenvalues between two zones, in the random matrix model corresponds to the
symmetry breaking $U(N)\to U(N-k)\times U(k)$, where $k$ is the number of eigenvalues in the second
zone, and is essentially nonperturbative. In the network case the symmetry breaking pattern could
be more subtle since we do not know explicitly the measure in the ensemble of adjacency matrices.
The eigenvalue tunnelling between two zones is fairly general phenomena in the matrix model
framework (see \cite{marino} for the review) meaning the formation of a kind of extended coherent
object, like a baby Universe. In our case, such tunnelling leads to the dense droplets (cliques)
formation, being the peculiar example of the global symmetry breaking.

%%%%%%%%%%%%%%%%%%%%%% Model %%%%%%%%%%%%%%%%%%%%%%%%%%%%%

The model under consideration is essentially the same as in \cite{valba}. We consider the canonical
ensemble of topological Erd\H{o}s-R\'enyi  graphs with quenched vertex degree. Each closed triple
of bonds (the closed triadic motif) is weighted with $\mu$. The partition function of the system is
\be
Z(\mu) = \sum_{\{\rm states\}} \hspace{-0.25cm} {\vphantom{\sum}}' e^{-\mu n_{\triangle}}
\label{eq:01}
\ee
where the prime in \eq{eq:01} means that the summation runs over all possible configurations of
nodes, under the condition of fixed degree $v_i$ in each vertex $i$ ($i=1,...,N$) of the graph. The
direct computation of $Z(\mu)$ in numerical simulations is very cumbersome, so, usually the numeric
algorithm runs as follows. The initial realization of ER network is prepared by connecting any
randomly taken pair of vertices with the probability $p$. Then, one randomly chooses two arbitrary
links, say, between vertices $i$ and $j$, $(ij)$ and between $k$ and $m$, $(km)$, and reconnect
them, getting new links $(ik)$ and $(jm)$. Such reconnection conserves the vertex degree
\cite{maslov}. Now one applies the standard Metropolis algorithm with the following rules: i) if
under the reconnection the number of close triples is increased, a move is accepted, ii) if the
number of close triples is decreased by $\Delta n_{\triangle}$, or remains unchanged, a move is
accepted with the probability $e^{-\mu \Delta n_{\triangle}}$. Then the Metropolis algorithm runs
repeatedly for large set of randomly chosen pairs of links, until it converges. In \cite{valba}
only the behavior of $\left<n_{\triangle}(\mu)\right>$ was considered. Here we turn to more
detailed investigation of the ground state in the constrained network.

%%%%%%%%%%%%%%%%%%%%%% Formulation of results %%%%%%%%%%%%%%%%%%%%%%%%%%%%%

\begin{figure}[ht]
\centerline{\includegraphics[width=6cm]{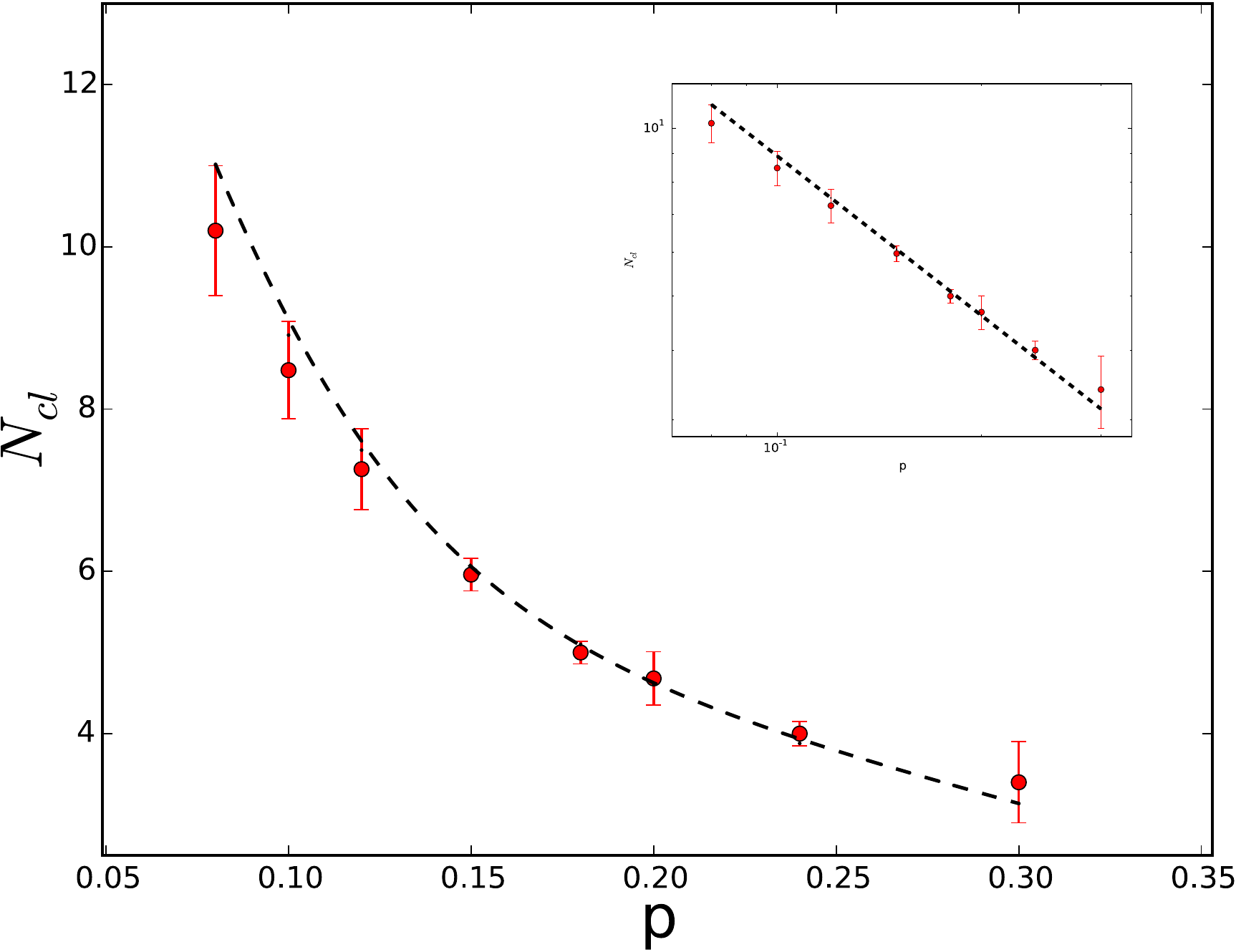}}
\caption{The number of clusters $N_{cl}$ as a function of the probability $p$ in ER graph.
The numerical data are obtained by averaging over 100 randomly generated graphs up to 512 vertices.
Numerical values are fitted by the curve $p^{-0.95}$; the behavior in doubly logarithmic scale is
shown in the insert.}
\label{fig01}
\end{figure}

We state that, given the probability, $p$, of the bond formation in the initial network, the
evolving network splits into the maximal number of clusters, $[p^{-1}]$ (where $[...]$ means the
integer part), being independent on the particular set of corresponding vertex degrees,
$\{v_1,...,v_N\}$ -- see the \fig{fig01}. In other words, the system with the maximal number of
clusters (cliques) is the proper ground state of the model. Thus, the system falls into the new
phase different from the Strauss one. Below we confirm this by analyzing the spectral density of
the adjacency matrix.

\begin{figure}[ht]
\centerline{\includegraphics[width=8cm]{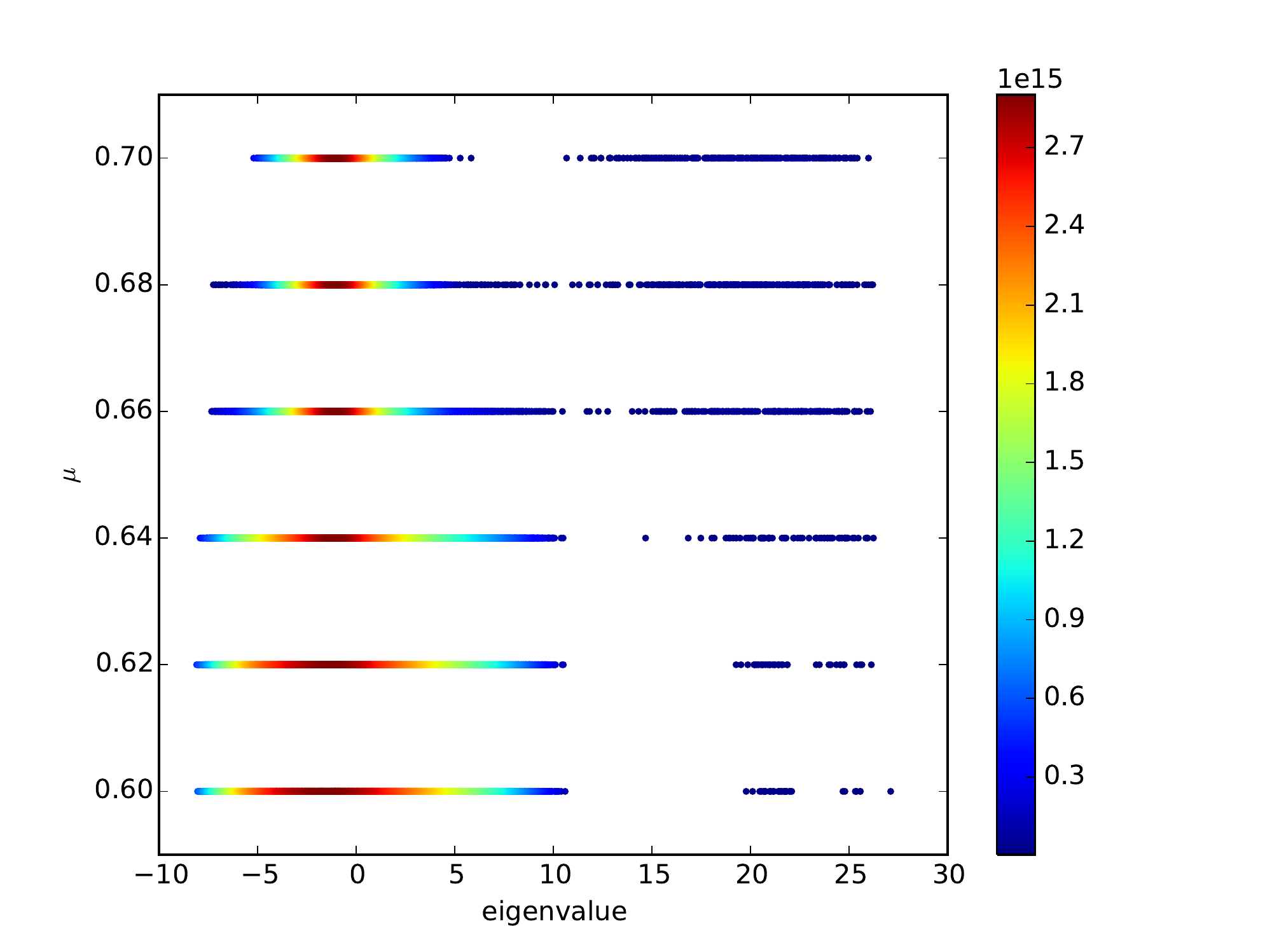}}
\caption{Spectral density of ensembles of ER graphs for different $\mu$. The numerical results are
obtained for 50 ER graphs of 256 vertices and $p=0.08$.}
\label{fig02}
\end{figure}

For any particular quenched network pattern and $\mu<\mu_c$, the spectral density has the shape
typical for ER graphs with moderate connection probability, $p=O(1)<1$, being the Wigner semicircle
with a single isolated eigenvalue apart. At $\mu_c$ the eigenvalues decouple from the main core and
a collection of isolated eigenvalues forms the second zone. The number of isolated eigenvalues
exactly coincides with the number of clusters formed above $\mu_c$. This perfectly fits with the
result of \cite{newman3}. Averaging over ensemble of graphs patterns smears the distribution of
isolated eigenvalues in the second zone. Above $\mu_c$ the support of SD in the first (central)
zone shrinks and the second zone becomes dense and connected -- see the \fig{fig02}.

\begin{figure}[ht]
\centerline{\includegraphics[width=8cm]{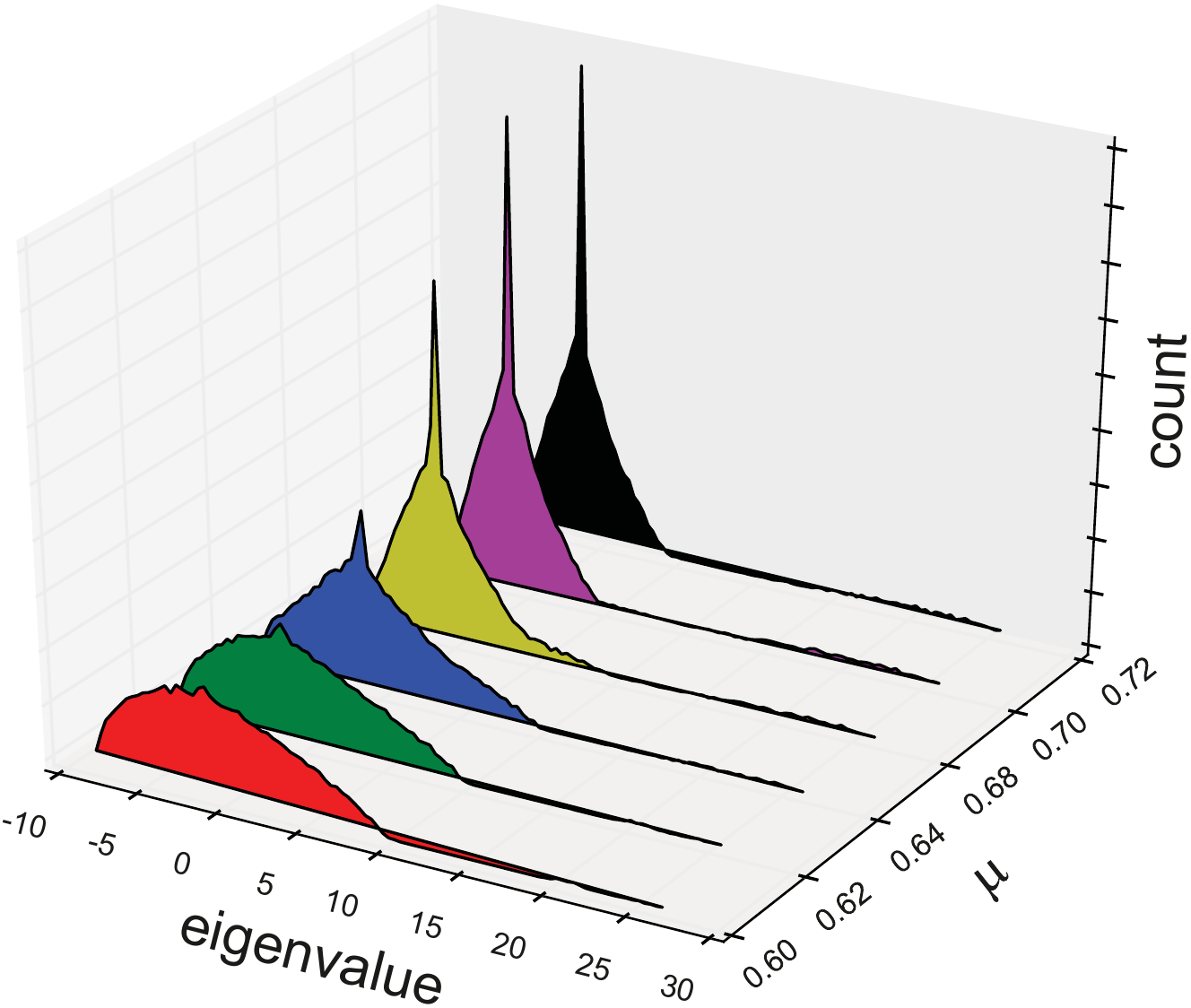}}
\caption{The spectral density in the ensembles of ER graphs for different $\mu$. The numerical
results are obtained for the ensembles of 50 random ER graphs of 256 vertices and the probability
$p=0.08$.}
\label{fig03}
\end{figure}

We have investigated SD inside each cluster (Cl) aiming to prove that Cl is almost full graph
(clique). We see that the shape of SD is drastically changed at $\mu_c$, see the \fig{fig03}, were
SD in the first zone below and above $\mu_c$ are shown. The triangle-shaped SD is typical for the
scale-free networks \cite{scale1,scale2}, however in our case the vertex degree is fixed at network
preparation and can be only redistributed between cliques. The remarkable point is that the SD
evaluated for each particular clique exhibits a hierarchical set of resonance peaks typical for
sparse matrices, which have been analytically investigated in \cite{av-kr-nech}.

\begin{figure}[ht]
\centerline{\includegraphics[width=8cm]{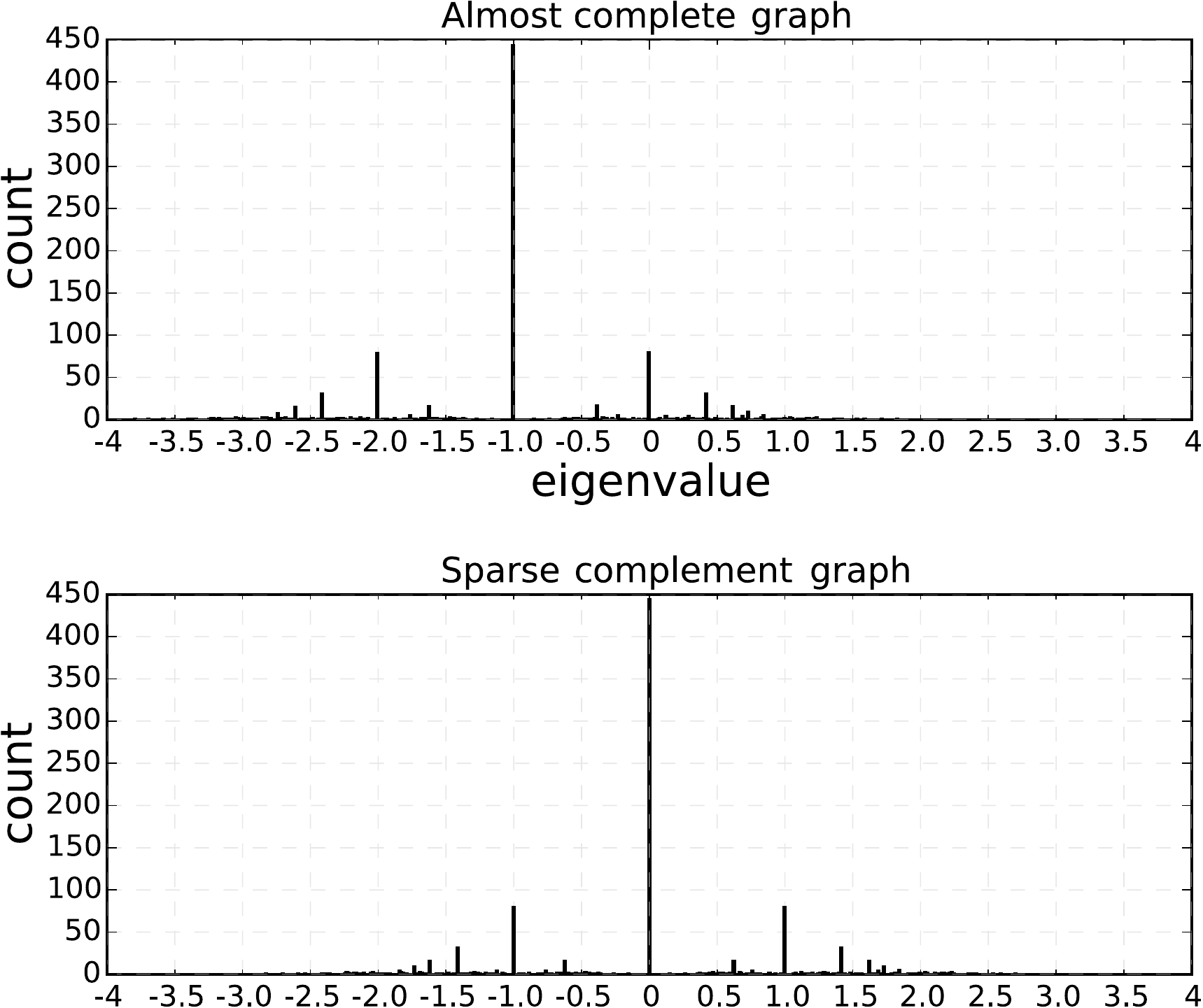}}
\caption{The duality between spectral densities of ensembles of almost fully connected graphs and
their sparse complements.}
\label{fig04}
\end{figure}

Using duality between the spectrum of sparse and almost full graphs \cite{tao}, we see from the
\fig{fig04} that SD in almost complete graph fits perfectly the shifted SD of sparse matrix
ensemble, meaning that our identification of clusters with cliques and separated eigenvalues is
true. The striking difference between the SD of single clique and the whole network indicates that
the triangle-shape SD of the whole networks occurs due to the inter-clique connections.

\begin{figure}[ht]
\centerline{\includegraphics[width=8.5cm]{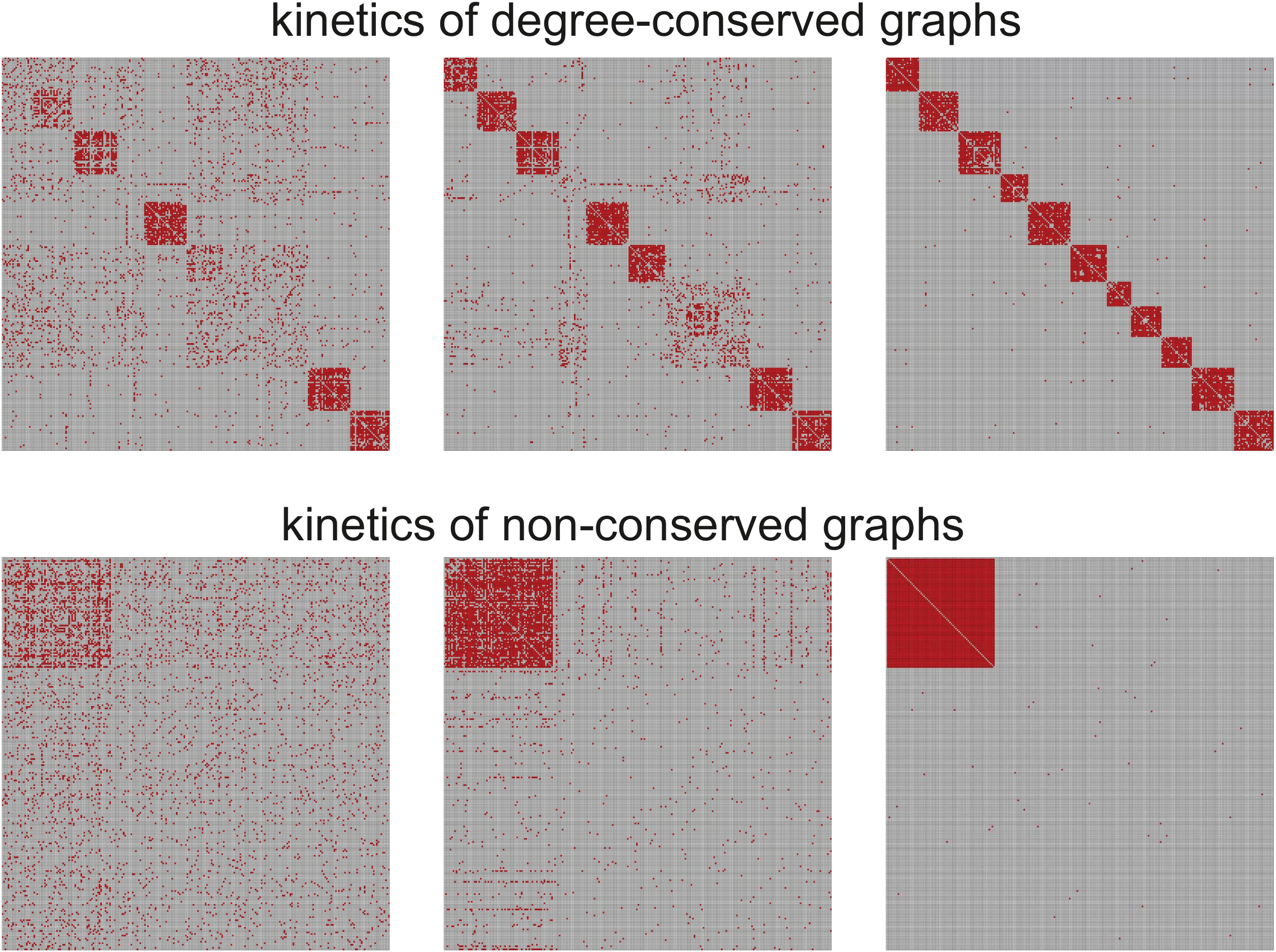}}
\caption{Few typical samples of intermediate stages of the network evolution: upper panel --
evolution with fixed vertex degree; lower panel -- eolution with non-fixed vertex degree.}
\label{fig05}
\end{figure}

It is very instructive to compare the typical adjacency matrices in the ground state of the ER
network with and without constraints at the same value of $p$. The corresponding matrices are
presented at \fig{fig05}, where different phases of the ground states are clearly seen. The ground
state in the Strauss phase consists of a single complete graph corresponding to the block in
adjacency matrix of some size, $k$, while the ground state of the quenched network involves
$[p^{-1}]$ almost complete graphs corresponding to blocks (cliques) in the adjacency matrix with
fluctuating sizes $N_i$ ($\sum_i N_i = N$) and the mean value in the clique,
$N_{cl}=\left<N_i\right>= N/[p^{-1}] \approx Np$.

To visualize the kinetics, we proceed as follows. First, we enumerate vertices at the preparation
condition in arbitrary order and run the Metropolis stochastic dynamics. When the system is
equilibrated and the cliques are formed, we re-enumerate vertices sequentially according to their
belongings to cliques. Then we restore corresponding dynamic pathways back to initial
configuration.

In the sparse regime there is the percolation phase transition in the ER network at
$p_{perc}=N^{-1}$ corresponding to the critical behavior of the $q=1$ Potts model. Due to the
duality, one could expect the dual percolation phase transition in the clique happens at
$\bar{p}_{perc} = 1 - N_{cl}^{-1}$. The dual percolation corresponds to the creation of
dislocations in the cliques extended through the whole droplet. The right tail of the spectral
density in the first zone behaves as
\be
\rho(\lambda)\big|_{\lambda\to\lambda_{\max}} \sim e^{-c/\sqrt{\lambda_{max}-\lambda}}
\label{eq:02}
\ee
($c$ is some positive constant) and is, as shown in \cite{av-kr-nech} for sparse ensembles, a
manifestation of a Lifshitz tail for the Andersen localization.

%Besides, we see that the typical topological structure of triads dominating at $\lambda_{max}$ in
%$\rho(\lambda)$, is such that two vertices belong to one clique, and third vertex -- to some other
%clique. Such "isosceles" triads are typical for Efimov localized tree-particle states. Due to their
%deep relation to Berezinsky-Kosterliz-Thouless (BKT) singularity in the correlation length
%\cite{stephanov}, one could speculate also about relation of the universal behavior \eq{eq:02} with
%the BKT phase transitions.

%%%%%%%%%%%%%%%%%%%%%% Comparison with the random matrix model %%%%%%%%%%%%%%%%%%%%%%%%%%%%%%%

Let us interpret our results in terms of the random matrix models postponing the detailed analysis
for the separate publication. The unconstrained Strauss model was analyzed in this context in
\cite{burda}. The choice of the classical potential in the large-$N$ network matrix model is
straightforward: the quadratic term fixes the number of links in the network, while the cubic term
encounters triangles since ${\rm Tr} M^3=N_{triangles}$. Therefore the matrix model for the Strauss
model reads as \cite{burda}
\be
Z(t_k)= e^{F}=\int dM e^{g^{-1}V(M)},
\ee
where $V(M)= a{\rm Tr} M^2 + \mu {\rm Tr} M^3$ and the parameter $a=\log(p^{-1} -1)$ for
unconstrained ER network. The Strauss model exhibits the phase transition at the critical value of
the chemical potential for triangles and is exact counterpart of the phase transition known in the
matrix model description of the pure 2d quantum gravity \cite{review}. However one should take
care comparing the random matrix models and the random adjacency matrix since the measures
in  the matrix integrals are different.

There is Riemann surface associated with any  matrix integral  where spectral density
in defined on
\be
y^2 =(V'(x))^2 + f(x)
\ee
where $y$ and $x$ are complex variables. The function $y(x)$ is related to the matrix model
resolvent, $\omega(x)= \left<{\rm Tr}\frac{1}{M-x}\right>$, as follows
\be
y(x)= 2\omega(x) + V'(x)
\ee
For the potential $V= ax^2 + \mu x^3$ the Riemann surface has genus one. The coefficients of
polynomial $f(x)$ are fixed by the filling fractions, $N_i$, around extrema of the potential $V(x)$
($V_{eff}'(a_i)=0$). In the cubic case for the Strauss model the eigenvalues are distributed in two
zones and if $k$ eigenvalues are in the second zone, the adjacency matrix  of the ground state
involves the size $k$ block filled by ones.

Thus, the second zone always exists, however in the unconstrained case all, but one, eigenvalues
belong to the first (central) zone with the Wigner semicircle distribution. The situation in the
constraint-driven case is different since fixed number of eigenvalues pass from the first zone to
the second one through the gap. From the conventional matrix model viewpoint the question can be
formulated a follows: why the fixed number of eigenvalues are forced to tunnel between zones above
the critical value of $\mu$ in the constrained model? The eigenvalue tunnelling is
very general phenomena, first discussed in the context of 2D gravity in \cite{david}. The review of
the general place of the eigenvalue tunnelling in topological strings and the matrix models can be
found in \cite{marino}. Depending of the physics described by the matrix model it corresponds to
the account of FZZT branes in the noncritical string, baby Universe creation in the quantum
gravity, or the creation of D-brane domain wall in the supersymmetric gauge theory.

It is convenient to introduce the constraints into the matrix model via Lagrangian multiplies. They
yield the linear term in the action  ${\rm Tr}\Lambda X$ with $\Lambda_{ij}= \{z_1\delta_{1j},...,
z_N\delta_{Nj}\}$ and one has to integrate over the Lagrangian multipliers, $z_i$. That is we have
a very peculiar version of the two-matrix model. The situation resembles the symmetry breaking by
the Wilson loop observables in the matrix model framework. In our case, qualitatively the form of
the constraint "selects the representation" of the Wilson loop and fixes the symmetry breaking
pattern. We can easily recognize that the pattern of the symmetry breaking realized in numerical
simulations indeed, is consistent with the form of constraint. Therefore, qualitatively we see that
the matrix model is consistent with two results of the numerical simulations: presence of
multi-zonal support for the spectral density, and the formation of cliques, by the mechanism of
inter-zone eigenvalue tunnelling.

The transition from the Wigner semicircle to the triangle-like SD is known in the matrix model
description of the Dirac operator spectrum in QCD \cite{ver,zahed} and admits deep physics behind.
If we scan QCD at finite volume at energies, smaller then the Thouless energy, $E_T$, which divides
the "ergodic" and "diffusive" regions in generic systems, the Dirac operator considered as the
Hamiltonian in 4+1 space-time enjoys the semicircle spectrum. In this regime the spectrum is
evaluated via the instanton liquid chiral matrix model and is saturated by the constant modes. On
the other hand at $E\gg E_T$, the non-constant modes are important and the SD at small $\lambda$
reads as
\be
\rho(\lambda) = \rho(0) - c|\lambda|
\ee
where $c>0$ is defined by the exchange of two soft Goldstone modes between two coherent states with
scalar quantum numbers \cite{smilga}. Such soft modes are present due to the spontaneous chiral
symmetry breaking. This perfectly fits with our observation that the triangle-shape SD in the
multi-clique phase definitely is due to the links between pairs of cliques. The global symmetry is
broken in our model above the phase transition as well, hence we could expect the presence of soft
modes which would play the role of diffusons and could provide the required inter-clique
interaction.

It is worth mention that nowadays the large-$N$ matrix model is interpreted as the theory of the
open string tachyon on the $N$ unstable D0 or ZZ branes \cite{verlinde,seiberg}. The final state of
the evolution of the unstable system is the coherent state of the closed string modes or stable D
branes. We have clear counterpart of this phenomena in our network as formation of the highly
coherent states -- set of cliques. Its number is fixed by the model from the very beginning. The
system on the stable FZZT branes identified as the Kontsevich-like matrix model \cite{gaiotto,vafa}
seems to be relevant for the description of the multi-clique phase of our model.

%%%%%%%%%%%%%%%%%%%%%%%%%%%%%%%%%%%% Conclusion %%%%%%%%%%%%%%%%%%%%%%%%%%%%%%%

In this Letter we described a decay of the constrained topological network into multi-clique phase
above some critical value of the chemical potential for triads. The decay has been analyzed via
evolution of the spectral density of the adjacency matrix. The ground state of the system above the
transition point is identified as interacting multi-clique states. The eigenvalue tunnelling is the
key point in our problem. This issue is a general nonperturbative phenomenon in matrix models
describing formation of extended objects, and our finding could provide new insight on it. We
believe that our model sheds additional light on the formation of stable D-brane from unstable ones
connected by strings. The similar spectral analysis of the constrained multicolor network will be
presented in a forthcoming publication \cite{color2}.

The imposed constraint is not exotic, being typical for the chemical, biological and social
networks. The phenomena we found can be considered as the operational tool to split the network
into the optimal droplets of almost full subgraphs (cliques) for generic random networks. Varying
the constraints, the required design of the ground state in the quantum network with the proper
symmetry breaking pattern can be manufactured.
%Also our model can be considered as the device for
%preparation of the scale-free network. To some extend the cliques are the vertices in this
%self-organized effective scale-free model.

The important finding concerns the emergence of the second zone and the genus one Riemann surface.
It immediately suggests that we could search for the action of the modular group on the torus
modulus. Such spectral Riemann surface nowadays is familiar in many examples and usually the
filling of the A-cycle of the torus corresponds to the "perturbative" degrees of freedom while the
filling of the B-cycle corresponds to the "nonperturbative" solitonic degrees of freedom which can
be build from the large number of the perturbative ones. The monopole and gauge boson are familiar
example of this phenomena. In our case we see the similar situation: the main "perturbative" zone
corresponds to the small perturbations, while the second zone is filled by the nonperturbative
solitonic cliques. It would be interesting to develop the "particle-soliton" duality for the
network ground state from the modular properties of the spectral torus. We postpone this
challenging question for the separate study.

We conclude by mentioning the possible relation of random network with quenched vertex degree with
some known physical models. In the context of quantum gravity (see for review \cite{ambjorn,
bianconi, trugenberger} the following question can be posed. The model under consideration is
topological and does not involve the metric structure. Is it possible to interpret clustering as
the appearance of the effective metric, see \cite{kryukov} for the recent discussion. We could
conjecture that the phase transition considered corresponds to the transition from the topological
$\left<g_{\mu\nu}\right>=0$ phase to the geometrical phase $\left<g_{\mu\nu}\right>\neq 0$ of the
network. The geometric phase could be related with the polymer phase of the 2d quantum gravity.

Another application deals with the budding phenomena (formation of bubble-like vesicles due to
spontaneous curvature) in lipid membranes \cite{lipowsky}. If the membrane is liquid, the material
can be redistributed over the whole tissue and only one vesicle is typically formed. However, in
presence of quenched disorder in the membrane, the redistribution of the material over the whole
sample is blocked and the formation of multi-vesicle phase seems plausible.

We are grateful to D. Kryukov and A. Mironov for the useful discussions. A.G. thanks the SCGP at
Stony Brook University where the part of the paper has been done  for the hospitality and support
during the program "Geometry of the Quantum Hall State". The work of A.G. was supported by grant of
Russian Science Foundation 14-050-00150 for IITP.


\begin{thebibliography}{99}

\bibitem{dorog}
S. Dorogovtsev, A. Goltsev and J. Mendes "`Critical phenomena in complex networks"' arxiv:0705.0010

\bibitem{strauss} D. Strauss, On a General Class of Models for Interaction,
SIAM Rev. {\bf 28}, 513 (1986)

\bibitem{burda} Z. Burda, J. Jurkiewicz, and A. Krzywicki, Network transitivity
and matrix models, Phys. Rev. E {\bf 69}, 026106

\bibitem{newman} J. Park and M. E. J. Newman, Solution for the properties of a
clustered network, Phys. Rev. E {\bf 72}, 026136

\bibitem{valba} M.V. Tamm, A.B. Shkarin, V.A. Avetisov, O.V. Valba, and S.K. Nechaev,
Islands of Stability in Motif Distributions of Random Networks, Phys. Rev. Lett. {\bf 113}, 095701

\bibitem{color}
 V.~Avetisov, A.~Gorsky, S.~Nechaev and O.~Valba,
  ``Spontaneous Symmetry Breaking and Phase Coexistence in Two-Color Networks,''
  Phys.\ Rev.\ E {\bf 93}, no. 1, 012302 (2016)
  doi:10.1103/PhysRevE.93.012302
  [arXiv:1506.00205 [cond-mat.stat-mech]].

      \bibitem{dorog1}
    S.N. Dorogovtsev, A.V. Goltsev, J.F.F. Mendes, A.N. Samukhin
    "`Spectra of complex networks"'
Phys. Rev. E 68, 046109 (2003)

    \bibitem{newman2}
Raj Rao Nadakuditi, M. E. J. Newman "`Graph spectra and the detectability of community structure in
networks"' Phys. Rev. Lett. 108, 188701 (2012)
 arXiv:1205.1813

 \bibitem{newman3}
 Nadakuditi, Raj Rao; Newman, M.E. J.
"`Spectra of random graphs with arbitrary expected degrees"' Physical Review E, vol. 87, Issue 1,
id. 012803 arXiv:1208.1275

\bibitem{newman4}
Zhang,  Xiao; Nadakuditi,  Raj  Rao;  Newman,  M.  E.  J. "`Spectra  of  random  graphs  with
community  structure  and  arbitrary  degrees"' Physical  Review  E,  Volume  89,  Issue  4,
id.042816 arXiv:1310.0046

\bibitem{av-kr-nech} V. Avetisov1, P.L. Krapivsky, and S. Nechaev, Native ultrametricity of
sparse random ensembles, J. Phys. A: Math. Theor. 49 (2016) 035101 (25pp)
doi:10.1088/1751-8113/49/3/035101   [arXiv:1506.05037]

\bibitem{scale1}
Illes J. Farkas, Imre Derenyi, Albert-Laszlo Barabasi, Tamas Vicsek "`Spectra of "Real-World"
Graphs: Beyond the Semi-Circle Law"' Physical Review E 64, 026704:1-12 (2001)

\bibitem{scale2}
K.-I. Goh, B. Kahng, D. Kim "`Spectra and eigenvectors of scale-free networks"' Phys. Rev. E 64,
051903 (2001)

\bibitem{tao} T. Tao and V. Vu, Random matrices: universality of local eigenvalue statistics,
Acta Math., 206(1):127–204 (2011);


  \bibitem{review}
    P.~H.~Ginsparg and G.~W.~Moore,
  "Lectures on 2-D gravity and 2-D string theory",
  hep-th/9304011

\bibitem{david}
  F.~David,
  "Nonperturbative effects in matrix models and vacua of two-dimensional gravity",
  Phys.\ Lett.\ B {\bf 302}, 403 (1993)
  doi:10.1016/0370-2693(93)90417-G
  [hep-th/9212106].

\bibitem{marino}
 M.~Mariño,
  "Lectures on non-perturbative effects in large $N$ gauge theories, matrix models and strings",
  Fortsch.\ Phys.\  {\bf 62}, 455 (2014)
  doi:10.1002/prop.201400005
  [arXiv:1206.6272 [hep-th]].

\bibitem{maslov} S. Maslov and K. Sneppen, Specificity and Stability in Topology of Protein
Networks, Science, {\bf 296}, 910 (2002)

      \bibitem{ver}
     J.~C.~Osborn, D.~Toublan and J.~J.~M.~Verbaarschot,
  From chiral random matrix theory to chiral perturbation theory",
  Nucl.\ Phys.\ B {\bf 540}, 317 (1999)
  doi:10.1016/S0550-3213(98)00716-0
  [hep-th/9806110].

    \bibitem{zahed}
    R.~A.~Janik, M.~A.~Nowak, G.~Papp and I.~Zahed,
  ``Chiral disorder in QCD,''
  Phys.\ Rev.\ Lett.\  {\bf 81}, 264 (1998)
  doi:10.1103/PhysRevLett.81.264
  [hep-ph/9803289].

     \bibitem{smilga}
  A.~V.~Smilga and J.~Stern,
  ``On the spectral density of Euclidean Dirac operator in QCD,''
  Phys.\ Lett.\ B {\bf 318}, 531 (1993).

\bibitem{verlinde}
      J.~McGreevy and H.~L.~Verlinde,
  ``Strings from tachyons: The c=1 matrix reloaded,''
  JHEP {\bf 0312}, 054 (2003)
  doi:10.1088/1126-6708/2003/12/054
  [hep-th/0304224].

    \bibitem{seiberg}
     I.~R.~Klebanov, J.~M.~Maldacena and N.~Seiberg,
  ``D-brane decay in two-dimensional string theory,''
  JHEP {\bf 0307}, 045 (2003)
  doi:10.1088/1126-6708/2003/07/045
  [hep-th/0305159].

    \bibitem{gaiotto}
     D.~Gaiotto and L.~Rastelli,
  ``A Paradigm of open / closed duality: Liouville D-branes and the Kontsevich model,''
  JHEP {\bf 0507}, 053 (2005)
  doi:10.1088/1126-6708/2005/07/053
  [hep-th/0312196].

    \bibitem{vafa}
    M.~Aganagic, R.~Dijkgraaf, A.~Klemm, M.~Marino and C.~Vafa,
  ``Topological strings and integrable hierarchies,''
  Commun.\ Math.\ Phys.\  {\bf 261}, 451 (2006)
  doi:10.1007/s00220-005-1448-9
  [hep-th/0312085].

\bibitem{color2}
V. Avetisov, A. Gorsky, M. Hovhannisyan, S. Nechaev, M. Tamm and O. Valba "Critical behavior of
multicolor constrained topological networks via eigenvalue tunneling"' to appear

 \bibitem{ambjorn}
  J.~Ambjorn, R.~Loll, Y.~Watabiki, W.~Westra and S.~Zohren,
  "New aspects of two-dimensional quantum gravity"
  Acta Phys.\ Polon.\ B {\bf 40}, 3479 (2009)
  [arXiv:0911.4208 [hep-th]].

  \bibitem{bianconi}
  G.~Bianconi and C.~Rahmede,
  "Network geometry with flavor: from complexity to quantum geometry,"
  Phys.\ Rev.\ E {\bf 93}, no. 3, 032315 (2016)
  doi:10.1103/PhysRevE.93.032315
  [arXiv:1511.04539 [cond-mat.stat-mech]]

\bibitem{trugenberger}
  C.~A.~Trugenberger,
  "Quantum Gravity as an Information Network: Self-Organization of a 4D Universe",
  Phys.\ Rev.\ D {\bf 92}, 084014 (2015)
  doi:10.1103/PhysRevD.92.084014
  [arXiv:1501.01408 [hep-th]].

\bibitem{kryukov}
    D. Krioukov,
    "Clustering implies geometry in networks",
     Phys. Rev. Lett. 116, 208302 (2016)
     arXiv:1604.01575

\bibitem{lipowsky} R. Lipowsky, Budding of membranes induced by intramembrane domains, J. de
Physique II, {\bf 2}, 1825 (1992); F. Julicher, R. Lipowsky, Domain-Induced Budding of Vesicles,
Phys. Rev. Lett., {\bf 70} 2964 (1993)

\end{thebibliography}
\end{document}